\documentclass[journal]{IEEEtran}
\usepackage{cite}

\ifCLASSINFOpdf
   \usepackage[pdftex]{graphicx}
\else
\fi
\usepackage{amsmath}
\usepackage{array}

  \usepackage[caption=false,font=footnotesize]{subfig}
\usepackage{multirow,makecell}
\usepackage{amsthm}
\usepackage{amssymb}
\usepackage{color}

\begin{document}

\onecolumn

%
\title{On Discrete Age of Information of Status Updating System With General Packet Arrival Processes}
%
%
%
\author{Jixiang~Zhang, Han~Xu, Daming~Cao, Yinfei~Xu, Minghao~Chen, and~Chengyu~Lin
\thanks{J. Zhang, M. Chen, and C. Lin are with School of IoT Engineering, Wuxi Taihu University, Wuxi, China. Email: \{zhangjx, chenmh, lincy\}@wxu.edu.cn.}
\thanks{H. Xu was with School of Information Science and Engineering, Southeast University, Nanjing, China. Email: han\_xu@seu.edu.cn.}
\thanks{D. Cao was with School of Electrical and Information Engineering, Nanjing University of Information Science and Technology, Nanjing, China. Email: dmcao@nuist.edu.cn.}
\thanks{Y. Xu was with School of Information Science and Engineering, Southeast University, Nanjing, China. Email: yinfeixu@seu.edu.cn.}
}
\maketitle

\begin{abstract}
Characterizing age of information (AoI) in status updating system with general packet arrival process and service process has great significance. In previous work \cite{1}, we have derived the explicit expressions of discrete average AoI when packet service time $S$ follows an arbitrary distribution. While assuming that packet interarrival time $Y$ is arbitrarily distributed, this paper obtains the exact characterization of discrete AoI in both preemptive and non-preemptive cases, by deriving AoI's probability generation function (PGF). Particularly, for preemptive case we prove that AoI's PGF can be obtained even when $Y$ and $S$ both follow arbitrary distributions.     
\end{abstract}


%
\IEEEpeerreviewmaketitle

\newtheorem{Definition}{Definition}  
\newtheorem{Theorem}{Theorem}  
\newtheorem{Lemma}{Lemma}
\newtheorem{Corollary}{Corollary}
\newtheorem{Proposition}{Proposition}
\newtheorem{Remark}{Remark}

\section{Introduction}
%
%
%
%

The development of 5G communication technology has made lots of Internet of Things (IoT) applications come true. In IoT applications such as autonomous vehicles and smart factories, the controller requires real-time data to determine next executed actions. The freshness of perceived data is crucial for the stable operations of IoT systems. In paper \cite{2}, the authors proposed using status updating system as the model of real-time communication system, in which the packet transmission process is described using queueing theory. To characterize transmission timeliness, they defined age of information (AoI) process, which tracks the age evolutions of the latest obtained packet at the receiver. In recent years, the theoretical analysis of AoI and its application in various real-time communication systems have attracted widespread attention. In papers \cite{3,4}, the authors conducted detailed investigation into contributions of AoI in analysis and application aspects.

For point-to-point bufferless status updating system, the aim of current paper is obtaining the exact characterization of discrete AoI, under the condition that the packet interarrival time $Y$ follows an arbitrary distribution. We consider both cases where a newly arrived packet can and cannot preempt the packet in the server. In non-preemptive case, we derive explicit expression of average AoI by calculating AoI's probability generation function (PGF). For preemptive case, it is proved that the PGF of discrete AoI can also be obtained even if packet interarrival time $Y$ and packet service time $S$ both follow arbitrary distributions. Using queuing theory terminology, in non-preemptive case we derive AoI's PGF for G/Geo/1/1 queue modeled system, and in preemptive case the PGF of AoI is determined for G/G/1/1 queue modeled status updating system. We generalize the methods in papers \cite{1} and \cite{5,6,7}, i.e., constructing a multi-dimensional random process to describe AoI evolutions and deriving average AoI by calculating the PGF, to overcome the difficulties caused by extending the distribution of $Y$ and $S$ to arbitrary distributions.

When extending the distributions of packet interarrival time or packet service time, the majority of existing works considered mainly obtaining system's average AoI. For example, for a multi-source M/M/1 queueing model, in works \cite{8,9} the authors first derived average AoI of each source, then provided three approximate expressions for average AoI when exponential serivce time is generalized to an arbitrary random variable. Calculating average AoI in multi-source system was also discussed in \cite{10} and \cite{11}. In paper \cite{10}, the authors obtained the exact expression of each source's average Aol in two cases. In the first one, the packet service time follows an exponential distribution, while in the second case the packet service time follows a general distribution. In addition, for each source in the system, in \cite{11} the authors obtained the Moment Generating Function (MGF) of AoI and peak AoI. Articles that consider generalizing packet interarrival time are relatively few. In paper \cite{12}, the authors considered discrete AoI in a D/G/1/1 non-preemptive queue modeled system, in which new packets are generated periodically. They derived expressions for time-dependent AoI distribution, expected AoI, and time-average AoI. For system with FCFS D/G/1 queues, the authors of \cite{13} studied AoI distribution and devised a problem of minimizing the tail of AoI distribution function. Peak AoI of D/G/1 queue modeled system was analyzed in paper \cite{14}, in which the authors estimated the outage probability that the peak AoI exceeds a certain threshold. For non-preemptive and preemptive cases, in paper \cite{15} Soysal and Ulukus determined exact average AoI expression for systems with G/G/1/1 queue. They considered the most general case where packet interarrival time and packet service time both have arbitrary distributions.

The remaining part of the paper is organized as follows. In Section \ref{sec2}, we depict the system model and introduce age of information process. In Section \ref{sec3}, discrete AoI is first analyzed in preemptive G/G/1/1 queue case. We derive the explicit expression of AoI's PGF. For non-preemptive G/Geo/1/1 queue modeled system, discrete AoI's PGF is determined in Section \ref{sec4}. Assuming $Y$ and $S$ follow geometric distributions, by calculating the derivative of PGF at $z=1$, in Section \ref{sec5} we obtain several specific expressions of bufferless system's average AoI. Finally, we conclude the paper in Section \ref{sec6}.

\section{System Model and Problem Formulation}\label{sec2}

\begin{figure}[!t]
\centering
\includegraphics[width=2.2in]{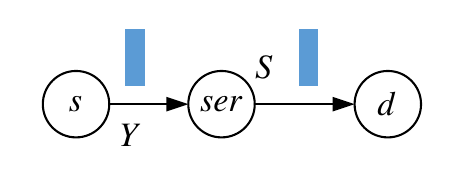}
\caption{Model of a discrete time status updating system with general packet arrival processes.}
\label{fig1}
\end{figure}

We depict a discrete time status updating system in Figure \ref{fig1}, in which random variables $Y$, $S$ denote the packet interarrival time and packet service time. 

The source $s$ generates new packets at random moments and the time interval between two consecutive packet generations is represented by $Y$. Through the transmitter these packets are sent from the source $s$ to the destination $d$. The process of transmitting packets is modeled as a queueing process, where the random propagation delay is denoted by service time $S$. We consider status updating systems without buffer. The preemptive case refers to the situation where newly arrived packets can preempt the packet that is currently in service. While in non-preemptive case, an arriving packet preempts the served one is not allowed.

Status updating system is the model of real-time communication systems, where the packets generated at the source are required to be transmitted to the destination as timely as possible. Age of information (AoI) is proposed to describe the evolutions of the latest packet's age in the receiver side. At the end of each time slot, if the receiver obtains a fresher packet, the value of AoI is updated to instantaneous age of this new packet. Otherwise, the value of AoI increases by 1 at the end of current time slot.

In recent years, average AoI has been applied widely as the metric to evaluate transmission timeliness of communication systems. Assuming that packet interarrival time $Y$ follows an arbitrary distribution, the aim of this paper is obtaining completed characterizition of discrete AoI for bufferless status updating systems. In Section \ref{sec3}, we first consider the preemptive case, while non-preemptive case is analyzed in Section \ref{sec4}.

\section{Discrete Age of Information of System with G/G/1/1 Queue: Preemptive Case}\label{sec3}

In preemptive case, in fact status updating systems with G/G/1/1 queue are considered, that is, we assume that both $Y$ and $S$ follow arbitrary distributions.  

Let $n$, $m$ be the AoI and age of packet in the server. Since there is no buffer, the age of served packet, $m$, is equal to the time that the served packet has experienced in the server. On the other hand, if packet preemption is allowed, then every new packet will enter the server directly and $m$ also tracks the time that has elapsed since last packet arrival. 
Constituting the random process $\{(n_k,m_k):n_k>m_k\geq0, k\in\mathbb N\}$, we show that the dynamics of state vector $(n,m)$ can be accurately described using the realizations of $Y$ and $S$. 

We determine random transfers of $(n,m)$, $n>m\geq1$ and $(n,0)$, $n\geq1$ as follows, which correspond to two different cases where the server is currently busy and currently idle. Denote $A$, $B$ as binary r.v.s indicating if the source generates a new packet and if the receiver obtains a fresher packet in a time slot. We use $A=1$ or $B=1$ to represent that the event occurs, and use $A=0$ or $B=0$ to denote that the event does not occur. According to the values of $A$ and $B$, all the random transfers and their transition probabilities are listed in Table \ref{table1}. 

\begin{table*}[!t]
\renewcommand{\arraystretch}{1.25}
\caption{Random Transfers of Two-dimensional State Vector in Preemptive Case}
\label{table1}
\centering
\scalebox{1}{
\begin{tabular}{c|c|l}
\Xhline{1pt}
\textbf{Initial state vector}  &  \textbf{Realization of $A$, $B$}  &  \textbf{Next state vector and transition probabilities}       \\
\Xhline{1pt}
\multirowcell{4} {$(n,m)$, $n>m\geq1$}   &  {$(A,B)=(0,0)$}  & {$(n+1,m+1)$ with prob. $P_{Y>m|Y>m-1}P_{S>m+1|S>m}$} \\
                                                                      &  {$(A,B)=(0,1)$}  & {$(m+1,0)$     with prob. $P_{Y>m|Y>m-1}P_{S=m+1|S>m}$} \\
                                                                      &  {$(A,B)=(1,0)$}  & {$(n+1,1)$      with prob. $P_{Y=m|Y>m-1}P_{S>1}$} \\
                                                                      &  {$(A,B)=(1,1)$}  & {$(1,0)$          with prob. $P_{Y=m|Y>m-1}P_{S=1}$} \\
\hline
\multirowcell{3} {$(n,0)$, $n\geq1$}  &  {$A=0$}               &  {$(n+1,0)$ with prob. $P_{Y>n|Y>n-1}$}\\
                                                               &  {$(A,B)=(1,0)$}   &  {$(n+1,1)$ with prob. $P_{Y=n|Y>n-1}P_{S>1}$}\\
                                                               &  {$(A,B)=(1,1)$}   &  {$(1,0)$     with prob. $P_{Y=n|Y>n-1}P_{S=1}$}\\
\Xhline{1pt}
\end{tabular}
}
\end{table*}

For initial state vector $(n,m)$, since the parameter $m$ tracks packet interarrival time and the experienced service time at the same time, we can obtain that 
\begin{align}
\Pr\{(n,m)&\to(n+1,m+1)\}=\Pr\{(A,B)=(0,0)\}  = P_{Y>m|Y>m-1}P_{S>m+1|S>m}  \label{E1}
\end{align}
and 
\begin{align}
\Pr\{(n,m)&\to(m+1,0)\}=\Pr\{(A,B)=(0,1)\}  = P_{Y>m|Y>m-1}P_{S=m+1|S>m}  \label{E2}
\end{align}
in which the conditional probabilities are used because the initial state vector $(n,m)$ has decided that $Y>m-1$ and $S>m$. Given the same $m$, the reason why $S$ is greater than $m$ but $Y$ is greater than $m-1$ is because we agree that whether the source generates a new packet is considered at the beginning of one time slot, and whether the packet service is completed is decided at the end of the time slot. The state vector $(n,0)$ actually implies that the source generated one packet in $n$ time slots ago and did not generate any other packets thereafter, and this packet has been transmitted to the receiver. Therefore, it shows that the packet interarrival time $Y$ must be greater than $n-1$. Except (\ref{E1}) and (\ref{E2}), probabilities that $(A,B)=(1,0)$ and $(A,B)=(1,1)$ can be determined accordingly. Based on above explanations about transition probabilities and the rule of AoI updatings, it is not hard to understand the random transfers in Table \ref{table1}.

For each $(n,m)$, in steady-state we define $\pi_{(n,m)}$ as its stationary probability. According to probability balance conditions in random process's steady-state, the stationary equations are established as follows.  
\begin{equation}\label{E3}
\begin{cases}
\pi_{(n,m)}=\pi_{(n-1,m-1)}P_{Y>m-1|Y>m-2}P_{S>m|S>m-1}   &    (n>m\geq2)   \\
\pi_{(n,1)}=\pi_{(n-1,0)}P_{Y=n-1|Y>n-2}P_{S>1}  + \sum\nolimits_{j=1}^{n-2}\pi_{(n-1,j)}P_{Y=j|Y>j-1}P_{S>1} & (n\geq3)  \\
\pi_{(2,1)}=\pi_{(1,0)}P_{Y=1}P_{S>1}       \\
\pi_{(n,0)}=\pi_{(n-1,0)}P_{Y>n-1|Y>n-2}+\sum\nolimits_{k=n}^{+\infty}\pi_{(k,n-1)}P_{Y>n-1|Y>n-2}P_{S=n|S>n-1}& (n\geq2) \\
\pi_{(1,0)}=\Big( \sum\nolimits_{n=1}^{+\infty}\pi_{(n,0)}P_{Y=n|Y>n-1}+ \sum\nolimits_{m=1}^{+\infty}\sum\nolimits_{n=m+1}^{+\infty}\pi_{(n,m)}P_{Y=m|Y>m-1} \Big)P_{S=1}
\end{cases}
\end{equation}

By referring to Table \ref{table1}, each equation in (\ref{E3}) is explained as how the state vector on the left side is obtained from the state vectors on the right side under corresponding conditions. From stationary equations (\ref{E3}), in this Section we derive the explicit expression of AoI's Probability Generation Function (PGF), which is defined to be 
\begin{equation}\label{E4}
H_{G/G}^P(z)=\sum\nolimits_{n=1}^{+\infty}z^n\Pr\{\Delta_{G/G}^P=n\}   
\end{equation}
where $\Delta_{G/G}^P$ is stationary AoI, $\Pr\{\Delta_{G/G}^P=n\}=\pi_{(n,0)}+\sum\nolimits_{m=1}^{n-1}\pi_{(n,m)}$ for $n\geq2$, while $\Pr\{\Delta_{G/G}^P=1\}=\pi_{(1,0)}$. 

Obviously, $H_{G/G}^P(z)$'s derivative at $z=1$ gives average AoI $\overline{\Delta}_{G/G}^P$. Substituting probability expressions gives that 
\begin{align}
H_{G/G}^P(z)&=\sum\nolimits_{n=1}^{+\infty}z^n\pi_{(n,0)} + \sum\nolimits_{m=1}^{+\infty}\sum\nolimits_{n=m+1}^{+\infty}z^n\pi_{(n,m)} = h_1(z) + h_2(z) \label{E5}
\end{align}
in which we denote that 
\begin{align}
h_1(z)&=\sum\nolimits_{n=1}^{+\infty}z^n\pi_{(n,0)}   \notag \\
h_2(z)&=\sum\nolimits_{m=1}^{+\infty}\sum\nolimits_{n=m+1}^{+\infty}z^n\pi_{(n,m)} \notag
\end{align}

\begin{Lemma}
Functions $h_1(z)$ and $h_2(z)$ in (\ref{E5}) are equal to
\begin{align}
h_1(z)&=\frac{1}{\mathbb E[Y]}\sum\nolimits_{n=1}^{+\infty}z^n P_{Y>n-1}P_{S\leq n}  \label{E6} \\
h_2(z)&=\frac{\left(\sum\nolimits_{n=1}^{+\infty}z^n P_{Y>n-1}P_{S> n}\right) \left(\sum\nolimits_{n=1}^{+\infty}z^n P_{Y=n}P_{S\leq n}\right)}{\mathbb E[Y]\left(1-\sum\nolimits_{n=1}^{+\infty}z^n P_{Y=n}P_{S>n}\right)}  \label{E7}
\end{align}
\end{Lemma}

\begin{IEEEproof}
Deriving closed-form expression of $H_{G/G}^P(z)$, equivalently $h_1(z)$ and $h_2(z)$ relies on clever handling of stationary equations (\ref{E3}). First of all, notice that in the last equation of (\ref{E3}), the sums in the bracket denote the probability that the source generates a new packet, which in fact is the total probability formula expanded based on different state vectors. Thus, we have 
\begin{equation}\label{E8}
\pi_{(1,0)}=\Pr\{\text{a new packet is generated}\}P_{S=1}=\frac{P_{S=1}}{\mathbb E[Y]}
\end{equation}
since on average the probability that the source generates a packet equals the reciprocal of average packet interarrival time. 

From first two rows, we prove that for $n>m\geq1$, $n-m\geq2$, 
\begin{align}
\pi_{(n,m)}&=\pi_{(n-1,m-1)}P_{Y>m-1|Y>m-2}P_{S>m|S>m-1}  \notag \\
&=\pi_{(n-2,m-2)}P_{Y>m-2|Y>m-3}P_{S>m-1|S>m-2} \times P_{Y>m-1|Y>m-2}P_{S>m|S>m-1}  \notag \\
&=\pi_{(n-2,m-2)}P_{Y>m-1|Y>m-3}P_{S>m|S>m-2}  \notag \\
&\qquad \qquad \qquad \qquad \qquad \qquad \qquad \qquad \vdots  \notag \\
&=\pi_{(n-m+1,1)}P_{Y>m-1}P_{S>m|S>1}  \notag \\
&=\left(\pi_{(n-m,0)}P_{Y=n-m|Y>n-m-1}+\sum\nolimits_{j=1}^{n-m-1}\pi_{(n-m,j)}P_{Y=j|Y>j-1} \right)P_{S>1}\times P_{Y>m-1}P_{S>m|S>1}  \notag \\
&=\Big(\pi_{(n-m,0)}P_{Y=n-m|Y>n-m-1} + \sum\nolimits_{j=1}^{n-m-1}\pi_{(n-m,j)}P_{Y=j|Y>j-1} \Big) P_{Y>m-1}P_{S>m}  \label{E9} 
\end{align}
and for cases of $n=m+1$, $m\geq1$, we show that 
\begin{equation}
\pi_{(m+1,m)}=\pi_{(2,1)}P_{Y>m-1}P_{S>m|S>1}=\Big(\pi_{(1,0)}P_{Y=1}\Big)P_{Y>m-1}P_{S>m}   \label{E10}
\end{equation}

Equations (\ref{E9}) and (\ref{E10}) are then applied to deal with the summation in the forth equation. We prove that  
\begin{align}
&\sum\nolimits_{k=n}^{+\infty}\pi_{(k,n-1)}P_{Y>n-1|Y>n-2}P_{S=n|S>n-1}  \notag \\
={}&\left(\pi_{(n,n-1)}+\sum\nolimits_{k=n+1}^{+\infty}\pi_{(k,n-1)}\right) P_{Y>n-1|Y>n-2}P_{S=n|S>n-1}  \notag \\
={}&\bigg(\pi_{(1,0)}P_{Y=1}P_{Y>n-2}P_{S>n-1}  \notag \\
  {}& +\sum\nolimits_{k=n+1}^{+\infty}\left[\pi_{(k-n+1,0)}P_{Y=k-n+1|Y>k-n}+\sum\nolimits_{j=1}^{k-n}\pi_{(k-n+1,j)}P_{Y=j|Y>j-1} \right]P_{Y>n-2}P_{S>n-1} \bigg)P_{Y>n-1|Y>n-2}P_{S=n|S>n-1}  \notag \\
={}&\Big( \sum\nolimits_{k=1}^{+\infty}\pi_{(k,0)}P_{Y=k|Y>k-1} + \sum\nolimits_{m=1}^{+\infty}\sum\nolimits_{n=m+1}^{+\infty}\pi_{(n,m)}P_{Y=m|Y>m-1}\Big)P_{Y>n-1}P_{S=n}\notag \\
={}&\frac{1}{\mathbb E[Y]}P_{Y>n-1}P_{S=n}  \label{E11}
\end{align}
In obtaining the second to last line, we made substitutions $k-n+1=n'$, $j=m'$ and exchanged the summation order. Using equation (\ref{E8}), the last line (\ref{E11}) is obtained.  

Then, the forth equation in (\ref{E3}) is simplified to be 
\begin{equation}
\pi_{(n,0)}=\pi_{(n-1,0)}P_{Y>n-1|Y>n-2} + \frac{1}{\mathbb E[Y]}P_{Y>n-1}P_{S=n} \label{E12}  
\end{equation}
Do once iteration yields that 
\begin{align}
\pi_{(n,0)}&=\left(\pi_{(n-2,0)}P_{Y>n-2|Y>n-3} + \frac{1}{\mathbb E[Y]}P_{Y>n-2}P_{S=n-1}\right)P_{Y>n-1|Y>n-2} + \frac{1}{\mathbb E[Y]}P_{Y>n-1}P_{S=n}   \notag \\
&=\pi_{(n-2,0)}P_{Y>n-1|Y>n-3}+\frac{1}{\mathbb E[Y]}P_{Y>n-1}\left(P_{S=n-1}+P_{S=n}\right)  \notag 
\end{align}
Continuing the iterations, we obtain that 
\begin{align}
\pi_{(n,0)}&=\pi_{(1,0)}P_{Y>n-1}+\frac{1}{\mathbb E[Y]}P_{Y>n-1}\left(P_{S=2}+\cdots+P_{S=n}\right) \notag \\
&=\frac{1}{\mathbb E[Y]}P_{S=1}P_{Y>n-1}+\frac{1}{\mathbb E[Y]}P_{Y>n-1}\left(P_{S=2}+\cdots+P_{S=n}\right) \notag \\
&=\frac{1}{\mathbb E[Y]}P_{Y>n-1}P_{S\leq n}  \label{E13}
\end{align}

Using (\ref{E13}), the first function $h_1(z)$ in equation (\ref{E6}) is obtained directly.  

Applying equations (\ref{E9}) and (\ref{E10}) again, according to its definition, $h_2(z)$ is calculated as follows.  
\begin{align}
h_2(z)&=\sum\nolimits_{m=1}^{+\infty}\sum\nolimits_{n=m+1}^{+\infty}z^n\pi_{(n,m)}  \notag \\
&=\sum\nolimits_{m=1}^{+\infty}z^{m+1}\pi_{(m+1,m)} + \sum\nolimits_{m=1}^{+\infty}\sum\nolimits_{n=m+2}^{+\infty}z^n\pi_{(n,m)}  \notag \\
&=\sum\nolimits_{m=1}^{+\infty}z^{m+1}\left(\pi_{(1,0)}P_{Y=1}\right)P_{Y>m-1}P_{S>m} \notag \\
&\quad + \sum\nolimits_{m=1}^{+\infty}\sum\nolimits_{n=m+2}^{+\infty}z^n \left(\pi_{(n-m,0)}P_{Y=n-m|Y>n-m-1}+\sum\nolimits_{j=1}^{n-m-1}\pi_{(n-m,j)}P_{Y=j|Y>j-1}\right)P_{Y>m-1}P_{S>m} \notag \\
&=\sum\nolimits_{m=1}^{+\infty}z^m \Big( \sum\nolimits_{n'=1}^{+\infty}z^{n'}\pi_{(n',0)}P_{Y=n'|Y>n'-1} + \sum\nolimits_{m'=1}^{+\infty}\sum\nolimits_{n'=m'+1}^{+\infty}z^{n'}\pi_{(n',m')}P_{Y=m'|Y>m'-1}\Big)  P_{Y>m-1}P_{S>m} \notag \\
&=\sum\nolimits_{m=1}^{+\infty}z^m \Big( h_1^*(z)+h_2^*(z) \Big)  P_{Y>m-1}P_{S>m} \label{E14}
\end{align}
in which we define that 
\begin{align}
h_1^*(z)&=\sum\nolimits_{n=1}^{+\infty}z^n\pi_{(n,0)}P_{Y=n|Y>n-1} \label{E15} \\
h_2^*(z)&=\sum\nolimits_{m=1}^{+\infty}\sum\nolimits_{n=m+1}^{+\infty}z^n\pi_{(n,m)}P_{Y=m|Y>m-1}  \label{E16}
\end{align}

From equation (\ref{E13}), $h_1^*(z)$ is determined directly to be  
\begin{equation}
h_1^*(z)=\frac{1}{\mathbb E[Y]}\sum\nolimits_{n=1}^{+\infty}z^n P_{Y=n}P_{S\leq n}  \label{E17}
\end{equation}
and by substituting (\ref{E9}) and (\ref{E10}), we prove that 
\begin{equation}
h_2^*(z)=\sum\nolimits_{m=1}^{+\infty}z^m P_{Y=m}P_{S>m}\Big( h_1^*(z) + h_2^*(z) \Big)  \label{E18}
\end{equation}
from which we solve that 
\begin{equation}
h_2^*(z) = \frac{\sum\nolimits_{m=1}^{+\infty}z^m P_{Y=m}P_{S>m}}{1-\sum\nolimits_{m=1}^{+\infty}z^m P_{Y=m}P_{S>m}}h_1^*(z)  \label{E19}
\end{equation}

Finally, substituting (\ref{E17}) and (\ref{E19}) into (\ref{E14}), it shows that 
\begin{align}
h_2(z)&=\left(\sum\nolimits_{n=1}^{+\infty}z^n P_{Y>n-1}P_{S>n}\right)\Big(h_1^*(z)+h_2^*(z)\Big)   \notag \\
&=\left(\sum\nolimits_{n=1}^{+\infty}z^n P_{Y>n-1}P_{S>n}\right) \left(1+ \frac{\sum\nolimits_{n=1}^{+\infty}z^n P_{Y=n}P_{S>n}}{1-\sum\nolimits_{n=1}^{+\infty}z^n P_{Y=n}P_{S>n}}\right)h_1^*(z)  \notag \\
&=\frac{\sum\nolimits_{n=1}^{+\infty}z^n P_{Y>n-1}P_{S>n}}{1-\sum\nolimits_{n=1}^{+\infty}z^n P_{Y=n}P_{S>n}}h_1^*(z) \notag \\
&=\frac{\left(\sum\nolimits_{n=1}^{+\infty}z^n P_{Y>n-1}P_{S>n}\right)\left(\sum\nolimits_{n=1}^{+\infty}z^n P_{Y=n}P_{S\leq n}\right)}{\mathbb E[Y]\left(1-\sum\nolimits_{n=1}^{+\infty}z^n P_{Y=n}P_{S>n}\right)}  \notag 
\end{align}
This derives the second function $h_2(z)$ in equation (\ref{E7}).    
\end{IEEEproof}

\begin{Theorem}
For bufferless status updating system with preemptive G/G/1/1 queue, the PGF of discrete AoI, $H_{G/G}^P$, is determined to be  
\begin{align}
H_{G/G}^P(z)= \frac{1}{\mathbb E[Y]}\sum\nolimits_{n=1}^{+\infty}z^n P_{Y>n-1}P_{S\leq n} + \frac{\left(\sum\nolimits_{n=1}^{+\infty}z^n P_{Y>n-1}P_{S> n}\right) \left(\sum\nolimits_{n=1}^{+\infty}z^n P_{Y=n}P_{S\leq n}\right)}{\mathbb E[Y]\left(1-\sum\nolimits_{n=1}^{+\infty}z^n P_{Y=n}P_{S>n}\right)}  \label{E20}
\end{align}
\end{Theorem}

\begin{IEEEproof}
From equation (\ref{E5}), $H_{G/G}^P(z)$ equals the sum of $h_1(z)$ and $h_2(z)$. 
\end{IEEEproof}

For specific distributions of $Y$ and $S$, according to (\ref{E20}) one can obtain AoI's PGF at the first place. Then, average AoI is determined by calculating PGF's derivative and substituting into $z=1$. Let $Y$ and $S$ be geometric r.v.s, in Section \ref{sec5} we will derive several explicit average AoI formulas, such as $\overline{\Delta}_{Ber/G}^P$ and $\overline{\Delta}_{G/Geo}^P$.

\section{Discrete Age of Information of System with G/Geo/1/1 Queue: Non-preemptive Case}\label{sec4}

For non-preemptive status updating system with Ber/G/1/1 queue, AoI's PGF and explicit average AoI formula have been obtained in work [1]. Here, let packet interarrival time $Y$ be an arbitrary random variable, we derive AoI's PGF and average AoI formula for G/Geo/1/1 queue modeled systems, in which it is assumed that packet service time $S$ follows geometric distribution $\Pr\{S=j\}=(1-\gamma)^{j-1}\gamma$, $j\geq1$.  

In non-preemptive case, although the age of served packet, $m$, still equals the service time that the packet has experienced, however, $m$ no longer tracks the elapsed time since last packet arrival. This is because every arriving packet resets the time interval since previous packet arrival to 0, but not every arriving packet resets the experienced service time to 0. To count the number of time slots that have passed since last packet arrival, we add an extra parameter $y$ in state vector, i.e., we use three-dimensional state vector $(n,m,y)$ and constitute random process $\{(n_k,m_k,y_k): k\in\mathbb N\}$.

Dividing into two cases where the server is currently busy and currently idle, which correspond to state vectors $(n,m,y)$, $n>m\geq y\geq0$ and $(n,0,y)$, $n>y\geq0$, respectively. Also using r.v.s $A$, $B$ to indicate in a time slot if the source generates new packets and whether the packet service is completed. According to realizations of $A$, $B$, we list all the state vector transfers and their transition probabilities in Table \ref{table2}. Assuming that the random process reaches the steady-state, we establish probability balance equations (\ref{E21}) as follows. 

\begin{table*}[!t]
\renewcommand{\arraystretch}{1.25}
\caption{Random Transfers of Three-dimensional State Vector in Non-Preemptive Case}
\label{table2}
\centering
\scalebox{1}{
\begin{tabular}{c|c|l}
\Xhline{1pt}
\textbf{Initial state vector}  &  \textbf{Realizations of $(A, B)$}  &  \textbf{Next state vector and transition probabilities}   \\
\Xhline{1pt}
\multirowcell{4} {$(n,m,y)$, \\ $n>m>y\geq1$}  & {$(A,B)=(0,0)$} & {$(n+1,m+1,y+1)$ with prob. $P_{Y>y+1|Y>y}(1-\gamma)$} \\
                    &  {$(A,B)=(0,1)$}  &    {$(m+1,0,y+1)$ with prob. $P_{Y>y+1|Y>y}\gamma$}  \\
                    &  {$(A,B)=(1,0)$}  &    {$(n+1,m+1,0)$ with prob. $P_{Y=y+1|Y>y}(1-\gamma)$}  \\
                    &  {$(A,B)=(1,1)$}  &    {$(m+1,0,0)$     with prob. $P_{Y=y+1|Y>y}\gamma$}  \\
\hline
\multirowcell{4} {$(n,m,0)$, \\ $n>m\geq1$}  & {$(A,B)=(0,0)$} & {$(n+1,m+1,1)$ with prob. $P_{Y>1}(1-\gamma)$} \\
                    &  {$(A,B)=(0,1)$}  &    {$(m+1,0,1)$     with prob. $P_{Y>1}\gamma$}  \\
                    &  {$(A,B)=(1,0)$}  &    {$(n+1,m+1,0)$ with prob. $P_{Y=1}(1-\gamma)$}  \\
                    &  {$(A,B)=(1,1)$}  &    {$(m+1,0,0)$     with prob. $P_{Y=1}\gamma$}  \\
\hline
\multirowcell{3} {$(n,0,y)$, \\ $n>y\geq1$}  & {$A=0$}  & {$(n+1,0,y+1)$ with prob. $P_{Y>y+1|Y>y}$}  \\
                    &  {$(A,B)=(1,0)$}  &    {$(n+1,1,0)$  with prob. $P_{Y=y+1|Y>y}(1-\gamma)$}  \\
                    &  {$(A,B)=(1,1)$}  &    {$(1,0,0)$      with prob. $P_{Y=y+1|Y>y}\gamma$}  \\
\hline
\multirowcell{3} {$(n,0,0)$, \\ $n\geq1$}  & {$A=0$}   & {$(n+1,0,1)$ with prob. $P_{Y>1}$}  \\
                    &  {$(A,B)=(1,0)$}  &    {$(n+1,1,0)$ with prob. $P_{Y=1}(1-\gamma)$}  \\
                    &  {$(A,B)=(1,1)$}  &    {$(1,0,0)$        with prob. $P_{Y=1}\gamma$}   \\
\Xhline{1pt}
\end{tabular}
}
\end{table*}

\begin{align}\label{E21}
\begin{cases}
\pi_{(n,m,y)}=\pi_{(n-1,m-1,y-1)}P_{Y>y|Y>y-1}(1-\gamma)   &  (n>m>y\geq2)  \\
\pi_{(n,m,1)}=\pi_{(n-1,m-1,0)}P_{Y>1}(1-\gamma)   &   (n>m\geq2)   \\  
\pi_{(n,m,0)}=\pi_{(n-1,m-1,0)}P_{Y=1}(1-\gamma) + \sum\nolimits_{j=1}^{m-2}\pi_{(n-1,m-1,j)}P_{Y=j+1|Y>j}(1-\gamma) & (n>m\geq3)   \\
\pi_{(n,2,0)}=\pi_{(n-1,1,0)}P_{Y=1}(1-\gamma)  &  (n\geq3)  \\
\pi_{(n,1,0)}=\pi_{(n-1,0,0)}P_{Y=1}(1-\gamma) + \sum\nolimits_{j=1}^{n-2}\pi_{(n-1,0,j)}P_{Y=j+1|Y>j}(1-\gamma) & (n\geq3) \\
\pi_{(2,1,0)}=\pi_{(1,0,0)}P_{Y=1}(1-\gamma) \\
\pi_{(n,0,y)}=\pi_{(n-1,0,y-1)}P_{Y>y|Y>y-1} + \sum\nolimits_{k=n}^{+\infty}\pi_{(k,n-1,y-1)}P_{Y>y|Y>y-1}\gamma & (n>y\geq2) \\
\pi_{(n,0,1)}=\pi_{(n-1,0,0)}P_{Y>1}+\sum\nolimits_{k=n}^{+\infty}\pi_{(k,n-1,0)}P_{Y>1}\gamma & (n\geq2)  \\
\pi_{(n,0,0)}=\sum\nolimits_{k=n}^{+\infty}\pi_{(k,n-1,0)}P_{Y=1}\gamma + \sum\nolimits_{j=1}^{n-2}\sum\nolimits_{k=n}^{+\infty}\pi_{(k,n-1,j)}P_{Y=j+1|Y>j}\gamma  &  (n\geq3)  \\
\pi_{(2,0,0)}=\sum\nolimits_{k=2}^{+\infty}\pi_{(k,1,0)}P_{Y=1}\gamma  \\ 
\pi_{(1,0,0)}=\sum\nolimits_{n=1}^{+\infty}\pi_{(n,0,0)}P_{Y=1}\gamma + \sum\nolimits_{y=1}^{+\infty}\sum\nolimits_{n=y+1}^{+\infty}\pi_{(n,0,y)}P_{Y=y+1|Y>y}\gamma  
\end{cases}
\end{align}

Explanations about state vector transfers and stationary equations (\ref{E21}) are similar to those given in Section \ref{sec3}, thus are omitted in this Section. In following paragraphs, we derive explicit formula of average AoI by calculating AoI's PGF.  

Define the PGF for non-preemptive case as 
\begin{align}
H_{G/Geo}(z)&=\sum\nolimits_{n=1}^{+\infty}z^n\Pr\{\Delta_{G/Geo}=n\}   \label{E22} \\
&=\sum\nolimits_{n=1}^{+\infty}z^n\pi_{(n,0,0)}+\sum\nolimits_{y=1}^{+\infty}\sum\nolimits_{n=y+1}^{+\infty}z^n\pi_{(n,0,y)} \notag \\
&\quad+\sum\nolimits_{m=1}^{+\infty}\sum\nolimits_{n=m+1}^{+\infty}z^n\pi_{(n,m,0)} + \sum\nolimits_{y=1}^{+\infty}\sum\nolimits_{m=y+1}^{+\infty}\sum\nolimits_{n=m+1}^{+\infty}z^n\pi_{(n,m,y)} \notag \\
&=h_1(z) + h_2(z) + h_3(z) + h_4(z) \label{E23}
\end{align}
that is, 
\begin{align}
h_1(z)&=\sum\nolimits_{n=1}^{+\infty}z^n\pi_{(n,0,0)}  \notag \\
h_2(z)&=\sum\nolimits_{y=1}^{+\infty}\sum\nolimits_{n=y+1}^{+\infty}z^n\pi_{(n,0,y)}  \notag \\
h_3(z)&=\sum\nolimits_{m=1}^{+\infty}\sum\nolimits_{n=m+1}^{+\infty}z^n\pi_{(n,m,0)}   \notag \\
h_4(z)&=\sum\nolimits_{y=1}^{+\infty}\sum\nolimits_{m=y+1}^{+\infty}\sum\nolimits_{n=m+1}^{+\infty}z^n\pi_{(n,m,y)} \notag 
\end{align}
and we explain that equation (\ref{E23}) is obtained by substituting probability expressions 
\begin{equation}\notag  
\Pr\{\Delta_{G/Geo}=n\}=\begin{cases}
\pi_{(1,0,0)},                                                &     (n=1)   \\
\pi_{(2,0,0)}+\pi_{(2,0,1)}+\pi_{(2,1,0)},  &     (n=2)   \\
\pi_{(n,0,0)}+\sum\nolimits_{y=1}^{n-1}\pi_{(n,0,y)}+\sum\nolimits_{m=1}^{n-1}\pi_{(n,m,0)}   &   (n\geq3)
\end{cases}
\end{equation}

\begin{Lemma}
We prove that the functions defined in (\ref{E23}) satisfy 
\begin{align}
h_1(z)&=\frac{[1-Y(1-\gamma)]\gamma z}{\mathbb E[Y]\left(1-Y[(1-\gamma)z]\right)}  \label{E24} \\
\left(1-Y[(1-\gamma)z]\right)h_2(z)&=(1-\gamma)Y(z)h_1(z) + \left((1-\gamma)Y(z)-Y[(1-\gamma)z]\right)h_2^{(m)}(z)   \label{E25} \\
h_3(z)&=\frac{z-Y(z)}{1-z}h_1(z) + \left(\frac{1-Y(z)}{1-z}-\frac{1-Y[(1-\gamma)z]}{1-(1-\gamma)z}\right) h_2^{(m)}(z) \label{E26}  \\
h_4(z)&=\frac{(1-\gamma)z-Y[(1-\gamma)z]}{1-(1-\gamma)z}h_2(z)  \label{E27}
\end{align}
in which $Y(z)=\sum\nolimits_{j=1}^{+\infty}z^n\Pr\{Y=j\}$ is the PGF of $Y$, and $h_2^{(m)}(z)=\sum\nolimits_{m=1}^{+\infty}\sum\nolimits_{n=m+1}^{+\infty}z^m\pi_{(n,m,0)}$ is determined as 
\begin{align}
h_2^{(m)}(z)=\frac{[1-Y(1-\gamma)](1-\gamma)z}{\mathbb E[Y]\left(1-Y[(1-\gamma)z]\right)}  \label{E28}
\end{align}
\end{Lemma}

\begin{IEEEproof}
Firstly, from the first two rows of (\ref{E21}), we have 
\begin{align}
\pi_{(n,m,y)}&=\pi_{(n-1,m-1,y-1)}P_{Y>y|Y>y-1}(1-\gamma)   \notag \\
&=\pi_{(n-2,m-2,y-2)}P_{Y>y-1|Y>y-2}(1-\gamma) \times P_{Y>y|Y>y-1}(1-\gamma)  \notag \\
&=\pi_{(n-2,m-2,y-2)}P_{Y>y|Y>y-2}(1-\gamma)^2  \notag \\
&\qquad \qquad \qquad \qquad \qquad  \vdots  \notag \\
&=\pi_{(n-y+1,m-y+1,1)}P_{Y>y|Y>1}(1-\gamma)^{y-1}  \notag \\
&=\pi_{(n-y,m-y,0)}P_{Y>1}(1-\gamma) \times P_{Y>y|Y>1}(1-\gamma)^{y-1}  \notag \\
&=\pi_{(n-y,m-y,0)}P_{Y>y}(1-\gamma)^y  \label{E29}
\end{align}
which is valid for $n>m>y\geq1$.

Equation (\ref{E29}) is used to deal with the summations in stationary equation for probabilities $\pi_{(n,0,0)}$, $\pi_{(n,m,0)}$, and $\pi_{(n,0,y)}$. Specifically, in stationary equation for $\pi_{(n,0,0)}$, it shows that  
\begin{align}
&\sum\nolimits_{j=1}^{n-2}\sum\nolimits_{k=n}^{+\infty}\pi_{(k,n-1,j)}P_{Y=j+1|Y>j}\gamma \notag \\
={}&\sum\nolimits_{j=1}^{n-2}\sum\nolimits_{k=n}^{+\infty}\pi_{(k-j,n-1-j,0)}P_{Y>j}(1-\gamma)^j\times P_{Y=j+1|Y>j}\gamma \notag \\
={}&\sum\nolimits_{j=1}^{n-2}\sum\nolimits_{k=n}^{+\infty}\pi_{(k-j,n-1-j,0)}P_{Y=j+1}(1-\gamma)^j\gamma  \notag
\end{align}
thus, for $n\geq2$, the stationary equation for $\pi_{(n,0,0)}$ becomes 
\begin{align}
\pi_{(n,0,0)}&=\sum\nolimits_{k=n}^{+\infty}\pi_{(k,n-1,0)}P_{Y=1}\gamma + \sum\nolimits_{j=1}^{n-2}\sum\nolimits_{k=n}^{+\infty}\pi_{(k-j,n-1-j,0)}P_{Y=j+1}(1-\gamma)^j\gamma  \notag \\
&=\sum\nolimits_{j=0}^{n-2}\sum\nolimits_{k=n}^{+\infty}\pi_{(k-j,n-1-j,0)}P_{Y=j+1}(1-\gamma)^j\gamma  \label{E30}
\end{align}

Similarly, in stationary equation of $\pi_{(n,m,0)}$, we calculate that 
\begin{align}
&\sum\nolimits_{j=1}^{m-2}\pi_{(n-1,m-1,j)}P_{Y=j+1|Y>j}(1-\gamma)  \notag \\
={}&\sum\nolimits_{j=1}^{m-2}\pi_{(n-1-j,m-1-j,0)}P_{Y>j}(1-\gamma)^j \times P_{Y=j+1|Y>j}(1-\gamma)  \notag \\
={}&\sum\nolimits_{j=1}^{m-2}\pi_{(n-1-j,m-1-j,0)}P_{Y=j+1}(1-\gamma)^{j+1}  \notag
\end{align}
then, it shows that for $n>m\geq3$, 
\begin{align}
\pi_{(n,m,0)}&=\pi_{(n-1,m-1,0)}P_{Y=1}(1-\gamma) + \sum\nolimits_{j=1}^{m-2}\pi_{(n-1-j,m-1-j,0)}P_{Y=j+1}(1-\gamma)^{j+1}  \notag \\
&=\sum\nolimits_{j=0}^{m-2}\pi_{(n-1-j,m-1-j,0)}P_{Y=j+1}(1-\gamma)^{j+1}  \label{E31}
\end{align}

Substituting (\ref{E29}) yields the recurrence relationships for $\pi_{(n,0,y)}$, and from which one can represent $\pi_{(n,0,y)}$ using expressions of $\pi_{(n,0,0)}$'s and $\pi_{(n,m,0)}$'s. For the sum in stationary equation of $\pi_{(n,0,y)}$, we have 
\begin{align}
&\sum\nolimits_{k=n}^{+\infty}\pi_{(k,n-1,y-1)}P_{Y>y|Y>y-1}\gamma    \notag \\
={}&\sum\nolimits_{k=n}^{+\infty}\pi_{(k-y+1,n-y,0)}P_{Y>y-1}(1-\gamma)^{y-1}\times P_{Y>y|Y>y-1}\gamma    \notag \\
={}&\sum\nolimits_{k=n-y+1}^{+\infty}\pi_{(k,n-y,0)}P_{Y>y}(1-\gamma)^{y-1}\gamma  \label{E32}
\end{align}
such that stationary equation for $\pi_{(n,0,y)}$ is transformed to be
\begin{equation}
\pi_{(n,0,y)}=\pi_{(n-1,0,y-1)}P_{Y>y|Y>y-1} + \sum\nolimits_{k=n-y+1}^{+\infty}\pi_{(k,n-y,0)}P_{Y>y}(1-\gamma)^{y-1}\gamma \label{E33}
\end{equation}

Using iteration equation (\ref{E33}) repeatedly, we show that   
\begin{align}
\pi_{(n,0,y)}&=\left(\pi_{(n-2,0,y-2)}P_{Y>y-1|Y>y-2} + \sum\nolimits_{k=n-y+1}^{+\infty}\pi_{(k,n-y,0)}P_{Y>y-1}(1-\gamma)^{y-2}\gamma \right) P_{Y>y|Y>y-1} \notag \\
&\quad + \sum\nolimits_{k=n-y+1}^{+\infty}\pi_{(k,n-y,0)}P_{Y>y}(1-\gamma)^{y-1}\gamma  \notag \\
&=\pi_{(n-2,0,y-2)}P_{Y>y|Y>y-2} + \gamma\sum\nolimits_{k=n-y+1}^{+\infty}\pi_{(k,n-y,0)}P_{Y>y}[(1-\gamma)^{y-2}+(1-\gamma)^{y-1}]  \notag \\
&\qquad \qquad \qquad \qquad \qquad \qquad \quad \vdots  \notag \\ 
&=\pi_{(n-y+1,0,1)}P_{Y>y|Y>1} + \gamma\sum\nolimits_{k=n-y+1}^{+\infty}\pi_{(k,n-y,0)}P_{Y>y}[(1-\gamma)+\cdots+(1-\gamma)^{y-1}]  \notag \\
&=\left(\pi_{(n-y,0,0)}P_{Y>1}+\sum\nolimits_{k=n-y+1}^{+\infty}\pi_{(k,n-y,0)}P_{Y>1}\gamma\right)P_{Y>y|Y>1}\notag \\
&\quad + \gamma\sum\nolimits_{k=n-y+1}^{+\infty}\pi_{(k,n-y,0)}P_{Y>y}[(1-\gamma)+\cdots+(1-\gamma)^{y-1}]  \notag \\
&=\pi_{(n-y,0,0)}P_{Y>y} + \gamma\sum\nolimits_{k=n-y+1}^{+\infty}\pi_{(k,n-y,0)}P_{Y>y}[1+(1-\gamma)+\cdots+(1-\gamma)^{y-1}]  \notag \\
&=\pi_{(n-y,0,0)}P_{Y>y} + \sum\nolimits_{k=n-y+1}^{+\infty}\pi_{(k,n-y,0)}P_{Y>y}[1-(1-\gamma)^y] \label{E34}
\end{align}

Now, the preparations for deriving functions $h_i(z)$, $i=1,2,3,4$ have been completed. According to their definition, each function is calculated. The obtained results are exact those we have listed in Lemma 2.  

First of all, for $h_1(z)$ it shows that 
\begin{align}
h_1(z)&=\sum\nolimits_{n=1}^{+\infty}z^n\pi_{(n,0,0)}=z\pi_{(1,0,0)} + \sum\nolimits_{n=2}^{+\infty}z^n\pi_{(n,0,0)}  \notag \\
&=z\left(\sum\nolimits_{n=1}^{+\infty}\pi_{(n,0,0)}P_{Y=1}\gamma + \sum\nolimits_{y=1}^{+\infty}\sum\nolimits_{n=y+1}^{+\infty}\pi_{(n,0,y)}P_{Y=y+1|Y>y}\gamma \right) \notag \\
&\quad + \sum\nolimits_{n=2}^{+\infty}z^n \sum\nolimits_{j=0}^{n-2} \sum\nolimits_{k=n}^{+\infty}\pi_{(k-j,n-1-j,0)}P_{Y=j+1}(1-\gamma)^j\gamma  \notag \\
&=z\left(M_1P_{Y=1}\gamma+\gamma M_1\sum\nolimits_{y=2}^{+\infty}P_{Y=y}+\gamma M_2\sum\nolimits_{y=1}^{+\infty}P_{Y=y+1}- \gamma M_2\sum\nolimits_{y=1}^{+\infty}P_{Y=y+1}(1-y)^y \right) + \frac{\gamma Y[(1-\gamma)z]}{1-\gamma}h_2^{(m)}(z) \notag \\
&=\left( \gamma M_1+\gamma M_2 - \frac{\gamma M_2 Y(1-\gamma)}{1-\gamma}\right)z + \frac{\gamma Y[(1-\gamma)z]}{1-\gamma}h_2^{(m)}(z)  \label{E35}
\end{align}
in which we denote $h_2^{(m)}(z)=\sum\nolimits_{m=1}^{+\infty}\sum\nolimits_{n=m+1}^{+\infty}z^m\pi_{(n,m)}$. The numbers $M_i$ is equal to $h_i(1)$ for $i=1,2,3,4$. That is, except $M_1$, $M_2$, we need also $M_3$ and $M_4$. 

Calculations to obtain (\ref{E35}) is not hard but tedious. We have to substitute $\pi_{(n,0,y)}$ using equation (\ref{E34}), which is long. The detailed derivations are omitted here, instead we have retained the result of each part in the second to last line. Readers can check these results themselves. 

Notice that equation (\ref{E35}) is different from (\ref{E24}) in Lemma 2, because there we have determined and substituted $M_1$, $M_2$, and $h_2^{(m)}(z)$. The other functions $h_2(z)$, $h_3(z)$, $h_4(z)$, also $h_2^{(m)}(z)$ are dealt with accordingly. In these derivations, equations (\ref{E29})-(\ref{E31}), and (\ref{E34}) are required to establish the equations of $h_i(z)$'s. For example, using (\ref{E34}) we have 
\begin{align}
h_3(z)&=\sum\nolimits_{y=1}^{+\infty}\sum\nolimits_{n=y+1}^{+\infty}z^n\pi_{(n,0,y)}  \notag \\
&=\sum\nolimits_{y=1}^{+\infty}\sum\nolimits_{n=y+1}^{+\infty}z^n\left(\pi_{(n-y,0,0)}P_{Y>y}+\sum\nolimits_{k=n-y+1}^{+\infty}\pi_{(k,n-y,0)}P_{Y>y}[1-(1-\gamma)^y]\right)  \notag \\
&=\sum\nolimits_{y=1}^{+\infty}\sum\nolimits_{n'=1}^{+\infty}z^{n'+y}\pi_{(n',0,0)}P_{Y>y} + \sum\nolimits_{y=1}^{+\infty}\sum\nolimits_{m'=1}^{+\infty}z^{m'+y}\sum\nolimits_{n'=m'+1}^{+\infty}\pi_{(n',m',0)}P_{Y>y}[1-(1-\gamma)^y] \notag \\
&=\sum\nolimits_{y=1}^{+\infty}\left(\sum\nolimits_{n'=1}^{+\infty}z^{n'}\pi_{(n',0,0)}\right) z^y P_{Y>y} 
+\sum\nolimits_{y=1}^{+\infty}\left(\sum\nolimits_{m'=1}^{+\infty}\sum\nolimits_{n'=m'+1}^{+\infty}z^{m'}\pi_{(n',m',0)}\right) P_{Y>y}\left(z^y-[(1-\gamma)z]^y\right)   \notag \\
&=h_1(z)\sum\nolimits_{y=1}^{+\infty}z^y\left(1- P_{Y\leq y}\right) 
+ h_2^{(m)}(z) \left( \sum\nolimits_{y=1}^{+\infty}z^y\left(1- P_{Y\leq y}\right)  -  \sum\nolimits_{y=1}^{+\infty}[(1-\gamma)z]^y \left(1- P_{Y\leq y}\right) \right)  \notag \\
&=h_1(z)\left(\frac{z}{1-z}-\frac{Y(z)}{1-z}\right) + h_2^{(m)}(z)\left(\frac{z-Y(z)}{1-z}-\frac{(1-\gamma)z-Y[(1-\gamma)z]}{1-(1-\gamma)z}\right)  \notag \\
&=\frac{z-Y(z)}{1-z}h_1(z) + \left(\frac{1-Y(z)}{1-z}-\frac{1-Y[(1-\gamma)z]}{1-(1-\gamma)z}\right) h_2^{(m)}(z)   \label{E36}
\end{align}
This derives equation (\ref{E26}) in Lemma 2. 

In equations (\ref{E24})-(\ref{E27}), letting $z=1$ obtains the relations between $M_i$'s. Combining with the condition $\sum\nolimits_{i=1}^4 M_i=1$, that is all of stationary probabilities add up to 1, these numbers can be solved explicitly. After $h_2^{(m)}(z)$ is also determined, then it can be proved that equation (\ref{E35}) is transformed to (\ref{E24}) in Lemma 2. 

So far, although not providing full calculation details, we have completed the proof of Lemma 2.  
\end{IEEEproof}

\begin{Theorem}
In non-preemptive case, the PGF of discrete AoI for G/Geo/1/1 queue modeled status updating system is determined to be  
\begin{equation}
H_{G/Geo}(z)=\frac{1-Y(z)}{1-z}\frac{[1-Y(1-\gamma)]\gamma z}{\mathbb E[Y]\left(1-Y[(1-\gamma)z]\right)[1-(1-\gamma)z]} \label{E37}
\end{equation}
\end{Theorem}

\begin{IEEEproof}
Using (\ref{E24})-(\ref{E28}) in Lemma 2, AoI's PGF for non-preemptive case, $H_{G/Geo}(z)$, is obtained as the sum of $h_i(z)$'s. 
\begin{align}
H_{G/Geo}(z)&=h_1(z) + h_2(z) + h_3(z) + h_4(z) \notag \\
&=h_1(z) + h_2(z) + \frac{z-Y(z)}{1-z}h_1(z) + \left(\frac{1-Y(z)}{1-z}-\frac{1-Y[(1-\gamma)z]}{1-(1-\gamma)z}\right)h_2^{(m)}(z) + \frac{(1-\gamma)z-Y[(1-\gamma)z]}{1-(1-\gamma)z}h_2(z)   \notag \\
&=\frac{1-Y(z)}{1-z}h_1(z) + \frac{1-Y[(1-\gamma)z]}{1-(1-\gamma)z}h_2(z) + \left(\frac{1-Y(z)}{1-z}-\frac{1-Y[(1-\gamma)z]}{1-(1-\gamma)z}\right)h_2^{(m)}(z)  \notag \\
&=\frac{1-Y(z)}{1-z}h_1(z) + \frac{(1-\gamma)Y(z)h_1(z)+\Big((1-\gamma)Y(z)-Y[(1-\gamma)z]\Big)h_2^{(m)}(z)}{1-(1-\gamma)z} \notag \\
&\quad + \left(\frac{1-Y(z)}{1-z} - \frac{1-Y[(1-\gamma)z]}{1-(1-\gamma)z}\right)h_2^{(m)}(z)  \notag \\
&=\frac{(1-z)+\gamma(z-Y(z))}{(1-z)[1-(1-\gamma)z]}h_1(z)+\frac{\gamma(z-Y(z))}{(1-z)[1-(1-\gamma)z]}h_2^{(m)}(z) \notag \\
&=\frac{(1-z)+\gamma(z-Y(z))}{(1-z)[1-(1-\gamma)z]}\frac{[1-Y(1-\gamma)]\gamma z}{\mathbb E[Y](1-Y[(1-\gamma)z])}+\frac{\gamma(z-Y(z))}{(1-z)[1-(1-\gamma)z]}\frac{[1-Y(1-\gamma)](1-\gamma)z}{\mathbb E[Y](1-Y[(1-\gamma)z])} \notag \\
&=\frac{[1-Y(z)][1-Y(1-\gamma)]\gamma z}{(1-z)\mathbb E[Y](1-Y[(1-\gamma)z])[1-(1-\gamma)z]} \label{E38}
\end{align}
this obtains PGF's expression (\ref{E37}) and we complete the proof.  
\end{IEEEproof}

\section{Average AoI in Several Special Cases}\label{sec5}

For preemptive and non-preemptive cases, we have obtained the PGFs of discrete AoI in Sections \ref{sec3} and \ref{sec4}. Let packet interarrival time $Y$ and packet service time $S$ follow geometric distributions, we first calculate the PGF $H_{G/G}^P(z)$ in (\ref{E20}) then obtain average AoI $\overline{\Delta}_{Ber/G}^P$ and $\overline{\Delta}_{G/Geo}^P$ by finding PGF's derivative at $z=1$. Using the PGF $H_{G/Geo}(z)$ in equation (\ref{E37}), average AoI $\overline{\Delta}_{G/Geo}$ in non-preemptive case is also determined.  

\begin{Corollary}
In preemptive case, for geometric $Y$ and arbitrarily distributed $S$, discrete AoI's PGF $H_{Ber/G}^P(z)$ and average AoI $\overline{\Delta}_{Ber/G}^P$ are obtained as 
\begin{align}
H_{Ber/G}^P(z)&=\frac{pS[(1-p)z]}{(1-p)(1-z)+pS[(1-p)z]}   \label{E39}   \\
\overline{\Delta}_{Ber/G}^P&=\frac{1-p}{pS(1-p)}    \label{E40}
\end{align}
in which $S(z)=\sum\nolimits_{j=1}^{+\infty}z^j\Pr\{S=j\}$ is the PGF of random variable $S$. While for arbitrarily distributed $Y$ and geometric $S$, the PGF $H_{G/Geo}^P(z)$ and average AoI $\overline{\Delta}_{G/Geo}^P$ are determined to be  
\begin{align}
H_{G/Geo}^P(z)&=\frac{\gamma z[1-Y(z)]}{\mathbb E[Y](1-z)[1-(1-\gamma)z]}   \label{E41}   \\
\overline{\Delta}_{G/Geo}^P&=\frac{1-\gamma}{\gamma}+\frac{\mathbb E[Y]+\mathbb E[Y^2]}{2\mathbb E[Y]}  \label{E42}
\end{align}
where $Y(z)=\sum\nolimits_{j=1}^{+\infty}z^j\Pr\{Y=j\}$ is the PGF of random variable $Y$.
\end{Corollary}

\begin{IEEEproof}
Let $Y$ or $S$ be a geometric random variable, firstly the general formula (\ref{E20}) is calculated, then average AoI is determined by deriving PGF's derivative at $z=1$. Equations (\ref{E39}) and (\ref{E40}) had been proved in work \cite{1}. We have proved that when $Y$ follows geometric distribution, for preemptive case the PGF $H_{G/G}^P(z)$ indeed degenerate to (\ref{E39}) and average AoI in (\ref{E40}) is obtained.  

For the case of arbitrary $Y$ and geometric $S$, the AoI's PGF (\ref{E41}) and average AoI formula (\ref{E42}) are determined as follows. 

We assume that $S$ has distribution $P_{S=j}=(1-\gamma)^{j-1}\gamma$ for $j\geq1$. Then, 
\begin{equation}
P_{S>n}=\sum\nolimits_{j=n+1}^{+\infty}P_{S=j}=\sum\nolimits_{j=n+1}^{+\infty}(1-\gamma)^{j-1}\gamma=(1-\gamma)^n \notag
\end{equation}
and $P_{S\leq n}=1-(1-\gamma)^n$. 

The four summations in (\ref{E20}) are calculated one by one. For the first one, 
\begin{align}
\text{Sum}_1&=\sum\nolimits_{n=1}^{+\infty}z^n P_{Y>n-1}P_{S\leq n}  \notag \\
&=\sum\nolimits_{n=1}^{+\infty}z^n P_{Y>n-1}[1-(1-\gamma)^n]  \notag \\
&=\sum\nolimits_{n=1}^{+\infty}z^n P_{Y>n-1} - \sum\nolimits_{n=1}^{+\infty}[(1-\gamma)z]^n P_{Y>n-1}   \notag \\
&=\sum\nolimits_{n=1}^{+\infty}z^n (1-P_{Y\leq n-1}) - \sum\nolimits_{n=1}^{+\infty}[(1-\gamma)z]^n (1-P_{Y\leq n-1}) \notag \\
&=\frac{z[1-Y(z)]}{1-z} - \frac{(1-\gamma)z\Big(1-Y[(1-\gamma)z]\Big)}{1-(1-\gamma)z}  \label{E43}
\end{align}

Then, for the second sum, we have 
\begin{align}
\text{Sum}_2&=\sum\nolimits_{n=1}^{+\infty}z^n P_{Y>n-1}P_{S> n}  \notag \\
&=\sum\nolimits_{n=1}^{+\infty}z^n P_{Y>n-1}(1-\gamma)^n  \notag \\
&=\sum\nolimits_{n=1}^{+\infty}[(1-\gamma)z]^n (1-P_{Y\leq n-1})  \notag \\
&=(1-\gamma)z\left(\sum\nolimits_{n=1}^{+\infty}[(1-\gamma)z]^{n-1} - \sum\nolimits_{n=1}^{+\infty}[(1-\gamma)z]^n (1-P_{Y\leq n}) \right)  \notag \\
&=\frac{(1-\gamma)z\Big(1-Y[(1-\gamma)z]\Big)}{1-(1-\gamma)z}  \label{E44}
\end{align}

We directly give the results of the remaining two sums as 
\begin{equation}
\text{Sum}_3=Y(z)-Y[(1-\gamma)z], \quad \text{Sum}_4=Y[(1-\gamma)z]  \label{E45}
\end{equation}

According to equation (\ref{E20}), omitting simple calculations we obtain the PGF $H_{G/Geo}^P(z)$ as  
\begin{align}
H_{G/Geo}^P(z)=\frac{1}{\mathbb E[Y]}\left(\text{Sum}_1+\frac{\text{Sum}_2\cdot\text{Sum}_3}{1-\text{Sum}_4}\right)
=\frac{[1-Y(z)]\gamma z}{\mathbb E[Y](1-z)[1-(1-\gamma)z]}  \label{E46}
\end{align}

Since average AoI is equal to PGF's derivative at $z=1$, thus we have 
\begin{equation}
\overline{\Delta}_{G/Geo}^P=\left.\frac{\mathrm d H_{G/Geo}^P(z)}{\mathrm d z}\right|_{z=1}
=\frac{\gamma}{\mathbb E[Y]}\left.\left(\frac{z[1-Y(z)]}{(1-z)[1-(1-\gamma)z]}\right)'\right|_{z=1}  \label{E47}
\end{equation}

Define
\begin{equation}
f(z)=\frac{z[1-Y(z)]}{(1-z)[1-(1-\gamma)z]}  \notag 
\end{equation}
then 
\begin{equation}
f(z)[1-(1-\gamma)z]=\frac{z[1-Y(z)]}{1-z}    \notag
\end{equation}

Taking the derivative in both sides yields that 
\begin{equation}
f'(z)[1-(1-\gamma)z] - (1-\gamma)f(z) = \frac{\Big(1-Y(z)-zY'(z)\Big)(1-z)+z[1-Y(z)]}{(1-z)^2}
\end{equation}
Substituting $z=1$ gives 
\begin{align}
f'(1)\gamma - \frac{(1-\gamma)\mathbb E[Y]}{\gamma}&=\lim_{z\to1}\frac{[1-Y(z)]-z(1-z)Y'(z)}{(1-z)^2}  \notag \\
&=\lim_{z\to1}\frac{-Y'(z)-(1-2z)Y'(z)-z(1-z)Y''(z)}{-2(1-z)}  \notag \\
&=\lim_{z\to1}\frac{2Y'(z)+zY''(z)}{2} \notag \\
&=\frac{2\mathbb E[Y]+\mathbb E[Y(Y-1)]}{2}  \notag \\
&=\frac{\mathbb E[Y]+\mathbb E[Y^2]}{2}  \label{E49}
\end{align}
In calculations, notice that $1=H_{G/Geo}^P(1)=\gamma f(1)/\mathbb E[Y]$, which implies that $f(1)=\mathbb E[Y]/\gamma$. 

From equation (\ref{E49}), we can solve that 
\begin{equation}
f'(1)\gamma = \frac{(1-\gamma)\mathbb E[Y]}{\gamma}+\frac{\mathbb E[Y]+\mathbb E[Y^2]}{2} \notag 
\end{equation}
thus
\begin{equation}
\overline{\Delta}_{G/Geo}^P=\frac{\gamma f'(1)}{\mathbb E[Y]}=\frac{1-\gamma}{\gamma}+ \frac{\mathbb E[Y]+\mathbb E[Y^2]}{2\mathbb E[Y]} \label{E50}
\end{equation}
this derives average AoI formula in (\ref{E42}), and the proof is completed. 
\end{IEEEproof}

\begin{Corollary}
For non-preemption case, average AoI of system with arbitrarily distributed $Y$, $\overline{\Delta}_{G/Geo}$, is obtained as 
\begin{equation}
\overline{\Delta}_{G/Geo}=\frac{(1-\gamma)Y'(1-\gamma)}{1-Y(1-\gamma)}+\frac{1}{\gamma}+\frac{\mathbb E[Y(Y-1)]}{2\mathbb E[Y]}    \label{Ez36}
\end{equation}
\end{Corollary}

\begin{IEEEproof}
Average AoI is obtained by calculating the derivative of (\ref{E37}), then substituting $z=1$. The calculations are similar to that of deriving $\overline{\Delta}_{G/Geo}^P$, thus are omitted.  
\end{IEEEproof}

All of average AoI formulas obtained above have been checked using special cases such as $\overline{\Delta}_{Ber/Geo}^P$ and $\overline{\Delta}_{Ber/Geo}$, which are average AoIs with geometric $Y$ and $S$. That is, we verify that 
\begin{equation}
\left.\overline{\Delta}_{G/Geo}^P\right|_{\text{geo}\, Y}=\overline{\Delta}_{Ber/Geo}^P,  \quad 
\left.\overline{\Delta}_{G/Geo}\right|_{\text{geo}\, Y}=\overline{\Delta}_{Ber/Geo}   \label{E52}
\end{equation}


\section{Conclusion}\label{sec6}

In this paper, we consider characterizing discrete AoI of bufferless status updating system with arbitrarily distributed packet interarrival time. For both preemptive and non-preemptive cases, the explicit expression of discrete AoI's PGF and average AoI formulas are obtained by the method of probability generation function. Particular, in preemptive case we derived the closed-form expression of AoI's PGF under the condition that packet interarrival time and packet service time both follow arbitrary distributions. We hope that the ideas and methods presented in this paper will be helpful in characterizing the AoI of other status updating systems.

\ifCLASSOPTIONcaptionsoff
  \newpage
\fi



\bibliographystyle{IEEEtran}
\bibliography{reff}
%
%

\end{document}